\documentstyle[prd,aps,preprint]{revtex}
\tightenlines
\input epsf.tex
\def\DESepsf(#1 width #2){\epsfxsize=#2 \epsfbox{#1}}

\begin{document}

\draft
\preprint{}
\title{Maximum And Minimum Dark Matter Detection Cross Sections} 

\author{  R. Arnowitt, B. Dutta and Y. Santoso }

\address{ Center For Theoretical Physics, Department of Physics, Texas A$\&$M
University, College Station TX 77843-4242}
\date{August, 2000}
\maketitle
\begin{abstract}
The range of neutralino-proton cross sections for R-parity preserving
supergravity models with GUT scale unification of the gauge coupling constants
is examined. The models considered here are mSUGRA, models with non universal
soft breaking and D-brane models. It is found that the current dark matter
detectors are sampling significant parts of the SUSY parameter space and future
detectors could sample almost the entire space. The special regions of
parameter space that may be inaccessible to future detectors are seen to have a
squark/gluino spectra beyond 1 TeV, but observable at the LHC.
\end{abstract}
\vspace{1.5cm} While the existence of dark matter, which is now believed to make
up about  30 $\%$ of the matter and energy of the universe, has been known for
sometime, the nature of dark matter remains unresolved. Supersymmetric models
with R parity invariance automatically posses a dark matter candidate, the
lightest supersymmetric particle (LSP). We consider here models based on gravity
mediated supergravity (SUGRA), where the LSP is generally the lightest
neutralino ($\tilde\chi^0_1$). Neutralinos in the halo of the Milky Way might be
directly detected by their scattering by terrestrial nuclear targets. Such
scattering has a spin independent and a spin dependent part. For heavy nuclear
targets the former dominates, and it is possible to extract then (to a good
approximation) the neutralino -proton cross section,
$\sigma_{\tilde\chi^0_1-p}$. Current experiments (DAMA, CDMS, UKMDC) have
sensitivity to halo $\tilde\chi^0_1$ for 
\begin{equation}
\sigma_{\tilde\chi^0_1-p}\stackrel{>}{\sim} 1\times 10^{-6} \, {\rm pb}
\end{equation} and future detectors (GENIUS, Cryoarray) plan to achieve a
sensitivity of 
\begin{equation}
\sigma_{\tilde\chi^0_1-p}\stackrel{>}{\sim} (10^{-9}-10^{-10}) \, {\rm pb}
\end{equation}

We consider here two questions: (1) what part of the parameter space is being
tested by current detectors, and (2) what is the smallest value of
$\sigma_{\tilde\chi^0_1-p}$ the theory is predicting  (i.e. how sensitive must
detectors be to cover the full SUSY parameter space). The answer to these
questions depends in part on the SUGRA model one is considering and also on what
range of theoretical and input parameter one assumes. In the following, we
examine three models that have been considered in the literature based on grand
unification of the gauge coupling constants at $M_G\cong 2\times 10^{16}$ GeV:
(1) Minimal super gravity GUT (mSUGRA) with universal soft breaking at $M_G$
\cite{SUGRA}; (2) Nonuniversal soft breaking models with Higgs and third
generation scalar masses at $M_G$ \cite{berezinsky}, and D-brane models (based
on type IIB orientifolds\cite{munoz}) which allow for nonuniversal gaugino
masses and nonuniversal scalar masses at $M_G$ \cite{brhlik}.

While each of the above models contain a number of unknown parameters, theories
of this type can still make relevant predictions for two reasons: (i) they allow
for radiative breaking of $SU(2)\times U(1)$ at the electroweak scale (giving a
natural explanation of the Higgs mechanism), and (ii) along with calculating
$\sigma_{\tilde\chi^0_1-p}$, the theory can calculate the relic density of
$\tilde\chi^0_1$, i.e
$\Omega_{\tilde\chi^0_1}=\rho_{\tilde\chi^0_1}/\rho_c$ where
$\rho_{\tilde\chi^0_1}$ is the relic mass density of  $\tilde\chi^0_1$ and
$\rho_c=3 H_0^2/8\pi G_N$ ($H_0$ is the Hubble constant and $G_N$ is the Newton
constant). Both of these greatly restrict the parameter space. In general one
has $\Omega_{\tilde\chi^0_1}h^2\sim(\int^{x_f}_0 dx\langle\sigma_{\rm
ann}v\rangle)^{-1}$ (where $\sigma_{\rm ann}$ is the neutralino annihilation
cross section in the early universe, $v$ is the relative velocity,
$x_f=kT_f/m_{\tilde\chi^0_1}$,
$T_f$ is the freeze out temperature, $\langle...\rangle$ means thermal average
and $h=H_0/100$ km s$^{-1}$Mpc$^{-1}$). The fact that these conditions can be
satisfied for reasonable parts of the SUSY parameter space represents a
significant success of the SUGRA models.

In the following we will assume $H_0=(70\pm 10)$km s$^{-1}$Mpc$^{-1}$ and matter
(m) and baryonic (b) relic densities of $\Omega_m=0.3\pm 0.1$ and
$\Omega_b=0.05$. Thus $\Omega_{\tilde\chi^0_1}h^2=0.12\pm 0.05$. The
calculations given below allow for a 2$\sigma$ spread, i.e. we take \cite{com} 
\begin{equation} 0.02\leq \Omega_{\tilde\chi^0_1}h^2\leq 0.25. 
\end{equation} It is clear that when the MAP and Planck sattelites determine the
cosmological parameters accurately, the SUGRA dark matter predictions will be
greatly sharpened.

\section{Calculational Details} In order to get reasonably accurate results, it
is necessary to include a number of theoretical corrections in the analysis. We
list here the main ones used in the calculations below: (i) In relating the
theory at $M_G$ to phenomena at the electroweak scale, the two loop gauge and
one loop Yukawa renormalization group equations (RGE) are used, iterating to get
a consistent SUSY spectrum. (ii) QCD RGE corrections are further included below
the SUSY breaking scale for contributions involving light quarks. (iii) A
careful analysis of the light Higgs mass $m_h$ is necessary (including two loop and
pole mass corrections) as the current LEP limits impact sensitively on the relic
density analysis for tan$\beta\leq 5$. (iv) L-R mixing terms are included in the
sfermion (mass)$^2$ matrices since they produce important effects for large
tan$\beta$ in the third generation. (v) One loop corrections are included to
$m_b$ and
$m_{\tau}$ which are again important for large tan$\beta$. (vi) The experimental
bounds on the $b\rightarrow s\gamma$ decay put significant constraints on the
SUSY parameter space and theoretical calculations here include the leading order
(LO) and approximate NLO corrections. We have not in the following imposed
$b-\tau$ (or $t-b-\tau$) Yukawa unification or proton decay constraints as these
depend sensitively on unknown post-GUT physics. For example, such constraints do
not naturally occur in the string models where
$SU(5)$ (or $SO(10)$) gauge symmetry is broken by Wilson lines at $M_G$ (even
though grand unification of the gauge coupling constants at $M_G$ for such
string models is still required).

All the above corrections are under theoretical control except for the
$b\rightarrow s\gamma$ analysis where a full NLO calculations has not been done.
(We expect that while the full analysis might modify the regions of parameter
space excluded by the $b\rightarrow s\gamma$ experimental constraint, the
minimum and maximum values of  $\sigma_{\tilde\chi^0_1-p}$ would probably not be
significantly changed.) The analysis of 
$\sigma_{\tilde\chi^0_1-p}$, taking into account the above theoretical
corrections has now been carried out by several groups obtaing results in
general agreement \cite{bottino,bottino1,falk,ellis,falk1,aads,aad}.  These
results are presented below.

Accelerator bounds significantly limit the SUSY parameter space. In the
following we assume the LEP bounds \cite{falk1} $m_h>104 (100)$ GeV for
tan$\beta$=3(5) and $m_{\chi^{\pm}_1}>104 (100)$ GeV. (For tan$\beta>5$, the
$m_h$ bounds do not produce a significant constraint\cite{fmw}.) For $b\rightarrow
s\gamma$ we assume an allowed range of $2\sigma$ from the CLEO data \cite{cleo}:
\begin{equation} 1.8\times 10^{-4}\leq B(B\rightarrow X_s\gamma)\leq 4.5 \times
10^{-4}
\end{equation} The Tevatron gives a bound of $m_{\tilde g}\geq 270$ GeV( for
$m_{\tilde q}\cong m_{\tilde g}$)\cite{comm1}.

Theory allows one to calculate the $\tilde\chi^0_1$-quark cross section and we
follow the analysis of \cite{ellis2} to convert this to $\tilde\chi^0_1-p$
scattering. For this one needs the $\pi-N$ $\sigma$ term,
\begin{equation}
\sigma _{\pi N}={1\over 2}(m_u+m_d)\langle p|{\bar u} u+{\bar d}d|p\rangle,
\end{equation}
$\sigma_0=\sigma _{\pi N}-(m_u+m_d)\langle p|{\bar s} s|p\rangle$ and the quark
mass ratio $r=m_s/{(1/2)(m_u+m_d)}$. We use here $\sigma _{\pi N}=65 $MeV, from
recent analyses \cite{mgo,pas} based on new $\pi-N$ scattering data,
$\sigma_0=30$ MeV\cite{bottino2} and r$=24.4\pm 1.5$\cite{leutwyler}. Older
$\pi-N$ data gave $\sigma _{\pi N}\cong 45$ GeV \cite{gs}, which if used would
reduce the overall
$\sigma_{\tilde\chi^0_1-p}$ by a factor of about 3.

\section{mSUGRA model} We consider first the mSUGRA model where the most
complete analysis has been done. mSUGRA depends on four parameters and one sign:
$m_0$ (universal scalar mass at $M_G$), $m_{1/2}$ (universal gaugino mass at
$M_G$), $A_0$ (universal cubic soft breaking mass, tan$\beta=\langle
H_2\rangle/{\langle H_1\rangle}$ (where $\langle H_{2,1}\rangle$ gives rise to
(up, down) quark masses) and
$\mu/|\mu|$( where $\mu$ is the Higgs mixing parameter in the superpotential,
$W_{\mu}=\mu H_1H_2$). One conventionally restricts the range of these
parameters by ``naturalness" conditions and in the following we assume $m_0\leq
1$ TeV, $m_{1/2}\leq 600$ GeV (corresponding to $m_{\tilde g}\leq 1.5$ TeV,
$m_{\tilde \chi^0_1}\leq 240$ GeV), $|A_0/m_0|\leq 5$, and 2$\leq tan\beta\leq$
50. Large tan$\beta$ is of interest since SO(10) models imply tan$\beta\geq 40$
and also
$\sigma_{\tilde\chi^0_1-p}$ increases with tan$\beta$.
$\sigma_{\tilde\chi^0_1-p}$ decreases with $m_{1/2}$ for large $m_{1/2}$, and
thus if one were to increase the bound on $m_{1/2}$ to 1 TeV ($m_{\tilde g}\leq
2.5$ TeV), the cross section would drop by a factor of 2-3.

The maximum $\sigma_{\tilde\chi^0_1-p}$ arise then for large tan$\beta$ and
small $m_{1/2}$. This can be seen in Fig.1 where
($\sigma_{\tilde\chi^0_1-p}$)$_{\rm max}$ is plotted vs. $m_{\tilde \chi^0_1}$
for tan$\beta$=20, 30, 40 and  50. Fig. 2 shows $\Omega_{\tilde\chi^0_1}h^2$ 
for tan$\beta=30$ when the cross section takes on  its maximum value. Current
detectors obeying Eq (1) are then sampling the parameter space for large tan
$\beta$, small $m_{\tilde \chi^0_1}$ and small $\Omega_{\tilde\chi^0_1}h^2$ i.e 
\begin{equation} tan\beta\stackrel{>}{\sim}25,\,m_{\tilde
\chi^0_1}\stackrel{<}{\sim} 90 {\rm
GeV},\,\Omega_{\tilde\chi^0_1}h^2\stackrel{<}{\sim} 0.1
\end{equation} 

To discuss the minimum cross section, it is convenient to
consider first 
$m_{\tilde \chi^0_1}\stackrel{<}{\sim} 150$ GeV ($m_{1/2}\leq 350$) where no
coannihilation occurs. The minimum cross section occurs for small tan$\beta$.
From Fig.3 one sees
\begin{equation}
\sigma_{\tilde\chi^0_1-p}\stackrel{>}{\sim} 4\times 10^{-9} {\rm pb};\,
 m_{\tilde \chi^0_1}\stackrel{<}{\sim} 140 {\rm GeV}
\end{equation} which would be accessible to detectors that are currently being
planned (e.g. GENIUS).

For larger $m_{\tilde \chi^0_1}$, i.e. $m_{1/2}\stackrel{>}{\sim} 150$ the
phenomena of coannihilation can occur in the relic density analysis since the
light stau, $\tilde \tau_1$, (and also $\tilde e_R$, $\tilde \mu_R$) can become
degenerate with the $\tilde\chi^0_1$. The relic density constraint can then be
satisfied in narrow corridor of $m_0$ of width 
$\Delta m_0\stackrel{<}{\sim}25$ GeV, the value of $m_0$ increasing as $m_{1/2}$
increases
\cite{falk}. Since $m_0$ and $m_{1/2}$ increase as one progresses up the
corridor, 
$\sigma_{\tilde\chi^0_1-p}$ will generally decrease.

We consider first the case $\mu>0$\cite{isa}. One finds in general that
$\sigma_{\tilde\chi^0_1-p}$ also decreases as $A_0$ increases. Fig.4 shows
$\sigma_{\tilde\chi^0_1-p}$ in the domain of large $A_0$ and for two values of
tan$\beta$. One sees that the smaller tan$\beta$ still gives  the lower cross
section, though the difference is mostly neutralized at larger $m_{1/2}$. (For
large tan$\beta$, $m_0$ also becomes large to satisfy the relic density
constraint i.e $m_0\cong 700$ GeV for tan$\beta$=40, $m_{1/2}=600$ GeV.) We have
in general for this regime
\begin{equation}
\sigma_{\tilde\chi^0_1-p}\stackrel{>}{\sim} 1\times 10^{-9} {\rm pb};\,{\rm
for}\,
 m_{1/2} \leq 600 {\rm GeV},\mu>0,\, A_0\leq4 m_{1/2}.
\end{equation}This is still within the sensitivity range of proposed detectors.

When $\mu$ is negative an ``accidental" cancellation can occur in part of the
parameter space in the coannihilation region which can greatly reduce
$\sigma_{\tilde\chi^0_1-p}$ \cite{ellis}. This can be seen in Fig.5, where starting with
small tan$\beta$ the cross section decreases, leading to a minimum at about
tan$\beta$=10, and then increases again for larger tan$\beta$. At the minimum
one has $\sigma_{\tilde\chi^0_1-p}\cong 1\times 10^{-12}$ for tan$\beta$=10 and
$m_{1/2}=600$ GeV. More generally one has \begin{equation}
\sigma_{\tilde\chi^0_1-p}< 1\times 10^{-10} {\rm pb}
\end{equation}for the parameter domain when $4\stackrel{<}{\sim}{\rm
tan}\beta\stackrel{<}{\sim}20$, $m_{1/2}\stackrel{>}{\sim} 450 {\rm GeV}
(m_{\tilde g}\stackrel{>}{\sim} 1.1 {\rm TeV})$, $\mu<0$. In this domain,
$\sigma_{\tilde\chi^0_1-p}$ would not be accessible to any of the  currently
planned detectors. However, mSUGRA also then predicts that this could happen
only when the gluino and squarks have masses greater than 1 TeV (and for only a
restricted region of tan$\beta$) a result that could be verified at the LHC.

\section{Nonuniversal SUGRA Models} In the discussion of SUGRA models with 
nonuniversal soft breaking, universality for the first two generations of
squark and slepton masses at $M_G$ is usually maintained to suppress flavor
changing neutral currents. One allows, however, the Higgs and third generation
squark and slepton masses to become nonuniversal. The masses may then be
parametrized at $M_G$ as follows:\begin{eqnarray} m_{H_{1}}^{\
2}&=&m_{0}^{2}(1+\delta_{1}); 
\quad m_{H_{2}}^{\ 2}=m_{0}^{2}(1+ \delta_{2});\\\nonumber m_{q_{L}}^{\
2}&=&m_{0}^{2}(1+\delta_{3}); \quad m_{t_{R}}^{\ 2}=m_{0}^{2}(1+\delta_{4});
\quad m_{\tau_{R}}^{\ 2}=m_{0}^{2}(1+\delta_{5}); 	\\\nonumber m_{b_{R}}^{\
2}&=&m_{0}^{2}(1+\delta_{6}); \quad m_{l_{L}}^{\ 2}=m_{0}^{2}(1+\delta_{7}).
\label{eq18}
\end{eqnarray} where $q_{L}\equiv (t_L, b_L)$, $l_{L}\equiv (\nu_\tau, \tau_L)$.
Here $m_0$ is the universal mass of the first two generations and $\delta_i$ are
the Higgs and third generation deviations from universality. In the following we
will restrict the $\delta_i$ to obey
\begin{equation} -1\leq\delta_i\leq1.\end{equation} The lower limit on
$\delta_i$ is necessary to prevent tachyonic behavior, and the condition
$\delta_i\leq 1$ is taken as a reasonable upper bound. We also maintain gauge
and gaugino mass unification at $M_G$.

While these models contain a large number of new parameters, their effects on 
$\sigma_{\tilde\chi^0_1-p}$ can be charcterized approximately by the signs of
$\delta_1...\delta_4$\cite{aads}. Thus the choice of
$\delta_3,\,\delta_4,\,\delta_1<0$ and
$\delta_2>0$ can greatly increase $\sigma_{\tilde\chi^0_1-p}$, and the reverse
choice can reduce $\sigma_{\tilde\chi^0_1-p}$ ( though by a much lesser amount).
The possible increase is shown in Fig.6 for tan$\beta=7$, $\mu>0$ where
nonuniversal soft breaking increases 
$\sigma_{\tilde\chi^0_1-p}$ by a factor of 10-100 compared to the universal
case. Thus it is possible for detectors to probe regions of smaller tan$\beta$
with nonuniversal breaking, and detectors obeying Eq. (1) can probe part of the
parameter space for tan$\beta$ as low as
$tan\beta\simeq 4$.

The minimum cross section occurs (as in mSUGRA) at the lowest tan$\beta$ and at
the largest
$m_{1/2}$ i.e. in the coannihilation region. We limit ourselves here to the
case where only the Higgs masses are nonuniversal i.e. $\delta_{1,2}\neq0$. One
finds then results similar to mSUGRA i.e.
$\sigma_{\tilde\chi^0_1-p}\stackrel{>}{\sim}10^{-9}$ pb for $\mu>0$, $m_{1/2}\leq$ 600
GeV. For $\mu<0$, there can again be a cancellation of matrix elements reducing
the cross section to $10^{-12}$ pb when $m_{1/2}= 600$ GeV in a restricted part
of the parameter space when tan$\beta\simeq$10\cite{comm2}.

\section{D-brane models} Recent advances in string theory have stimulated again
the construction of string inspired phenomenologically viable models. One type,
based on type IIB orientifolds \cite{munoz} puts the Standard Model on sets of 5
branes. These models are of interest in that they can lead to a pattern of soft
breaking different from what can naturally arise in conventional supergravity
GUTs. An interesting example\cite{brhlik} has the following soft breaking
pattern at $M_G$:
\begin{eqnarray}\tilde m_1&=&\tilde m_3=-A_0=\sqrt{3} cos\theta_b \Theta_1
e^{-i\alpha_1}m_{3/2}\\\nonumber
\tilde m_2&=&\sqrt{3} cos\theta_b (1-\Theta_1^2)^{1/2}m_{3/2}
\end{eqnarray} where and $\tilde m_i$ are the gaugino masses, and 
\begin{eqnarray} m_{12}^2&=&(1-3/2 sin^2\theta_b)m_{3/2}^2\, {\rm for}\,
q_L,\,l_L,\,H_1,\,H_2\\\nonumber m_{1}^2&=&(1-3 sin^2\theta_b)m_{3/2}^2\, {\rm
for}\, u_R,\,d_R,\, e_R.
\end{eqnarray} Thus the $SU(2)$ doublets are all degenerate at $M_G$ but
different from the singlets. We note Eq. (13) implies $\theta_b<$0.615.

We consider first the case with no CP violating phases in the soft breaking
sector, i.e.
$\alpha_1$=0 and the $\mu$ parameter real. In general
$\sigma_{\tilde\chi^0_1-p}$ is a decreasing function of $\theta_b$ since the
squark and slepton masses increase as $\theta_b$ decreases. This is illustrated
in Fig. 7 where $\sigma_{\tilde\chi^0_1-p}$ is plotted for tan$\beta$=10,
$m_{3/2}=175$  GeV with 
$\theta_b=0.5$ (upper curve) and $\theta_b=0.2$ (lower curve). One has that
current detectors obeying Eq. (1) are sensitive to the model in the domain when
tan$\beta\stackrel{>}{\sim}15$. However $m_{\tilde \chi^0_1}$ is quite small
when tan$\beta$ is close to the minimum value.

The lower bound on $\sigma_{\tilde\chi^0_1-p}$ should occur when $\theta_b$ and
tan$\beta$ are small. Fig. 8 shows the minimum cross section for
$m_{\tilde\chi^0_1}\leq 300$ GeV, $\mu>0$. One sees 
$\sigma_{\tilde\chi^0_1-p}\stackrel{>}{\sim}10^{-9}$ pb. As
in mSUGRA, a cancellation of matrix elements can occur for $\mu<0$, allowing a
smaller cross section in the domain
$7\stackrel{<}{\sim}$tan$\beta\stackrel{<}{\sim}30$ and reaching a minimum of
$\simeq10^{-12}$ pb for $m_{3/2}=600$ GeV at tan$\beta\simeq 12$. These low  cross sections again
imply very heavy gluinos i.e $m_{\tilde g}\stackrel{>}{\sim}950$ GeV. The D-brane
model also possesses an interesting new region of coannihilation for
$\Theta_1\simeq 0.8$ when the light chargino
$\tilde\chi^{\pm}_1$ can become degenerate with $\tilde\chi^0_1$, which has
not been analyzed here. 

CP violating phases produce contributions to the electric dipole moments of the
electron $d_e$, and neutron $d_n$. The bound on $d_e$ is $d_e<4.3\times
10^{-27}$ ecm  at 95 $\%$ C.L.\cite{de}, and one must constrain the parameter
space to satisfy this. This generally reduces
$\sigma_{\tilde\chi^0_1-p}$ by a factor of 2-3\cite{aad}. However models with CP
violating phases require a great deal of fine tuning unless tan$\beta\leq$ 3-5,
so one would expect in this case
$\sigma_{\tilde\chi^0_1-p}$ to be quite small.

\section{Summary} We have examined here the neutralino-proton cross section
for a number of SUGRA type models. In all the models considered, there are
regions of parameter space with
$\tilde\chi^0_1-p$ cross sections of the size that could be observed with
current detectors. Thus with the sensitivity of Eq. (1), detectors would be
sampling regions of the parameter space for mSUGRA where
tan$\beta\stackrel{>}{\sim}25$, $m_{\tilde \chi^0_1}\stackrel{<}{\sim}
 90 {\rm GeV}$ and $\Omega_{\tilde \chi^0_1}h^2\stackrel{<}{\sim}0.1$.
Nonuniversal models
 can have larger cross sections and so detectors could sample down to  
 tan$\beta\stackrel{>}{\sim}4$, while for the D-brane models considered,
detectors could sample
 down to tan$\beta\stackrel{>}{\sim}15$.
 
 The minimum cross sections these models predict are considerably below current
sensitivity.
 Thus for mSUGRA one finds for $\mu>0$ 
\begin{equation}
\sigma_{\tilde\chi^0_1-p}\stackrel{>}{\sim} 1\times 10^{-9} {\rm pb}\,\,{\rm for}
\, m_{1/2} \leq 600 {\rm GeV},\,\mu>0.
\end{equation}where $m_{1/2}=600$ GeV corresponds to $m_{\tilde g}\cong1.5$ TeV, 
$m_{\tilde \chi^0_1}\cong240$ TeV. This is still in the range that would be
accessible to detectors being planned (such as GENIUS or Cryoarray). For
$\mu<0$, a cancellation can occur in certain regions of parameter space allowing
the cross sections to fall below Eq. (14), Thus 
\begin{equation}
\sigma_{\tilde\chi^0_1-p}<1\times 10^{-10} {\rm pb}\,\,{\rm for}\,
4\stackrel{<}{\sim}{\rm tan}\beta\stackrel{<}{\sim}20,\,\mu<0,\,
m_{1/2}\stackrel{>}{\sim} 450 {\rm GeV}
\end{equation} and reaching a minimum of $\sigma_{\tilde\chi^0_1-p}\cong 1\times
10^{-12}$ pb for tan$\beta$=10, $m_{1/2}$=600 GeV, $\mu<0$. This domain would
appear not to be accessible to future planned detectors. Since  $m_{1/2}$=450
GeV corresponds to $m_{\tilde g}\cong1.1$ TeV, this region of parameter space
would imply a gluino squark spectrum at the LHC above 1 TeV. 

The above results
holds for the mSUGRA model. While a full analysis of coannihilation has not been
carried out for the nonuniversal and D-brane models, results similar to the
above hold for these over large regions of parameter space. Thus for
nonuniversal Higgs masses and for the
 D-brane model one finds $\sigma_{\tilde\chi^0_1-p}\stackrel{>}{\sim}10^{-9}$ pb for
$\mu>0$, while a cancellelation allows $\sigma_{\tilde\chi^0_1-p}$ to fall to
$10^{-12}$ pb for $\mu<0$ at tan$\beta\simeq 10$ (with again a gluino/squark mass spectrum
in the TeV domain).

In the above we have limited $m_{1/2}$ to $m_{1/2}\leq 600$ GeV ($m_{\tilde
g}\leq1.5$ TeV). Increasing $m_{1/2}$ lowers the cross section. Thus allowing 
$m_{1/2}= 1$ TeV ($m_{\tilde g}\cong2.5$ TeV) would reduce the minimum cross
section by a factor of about 2-3. Also using the earlier value of $\sigma_{\pi N}=45 $
MeV \cite{gs} rather than the most recent value $\sigma_{\pi N}=65 $ MeV
\cite{mgo,pas} would reduce $\sigma_{\tilde\chi^0_1-p}$ by a factor of 3.

\newpage
\begin{figure}[htb]
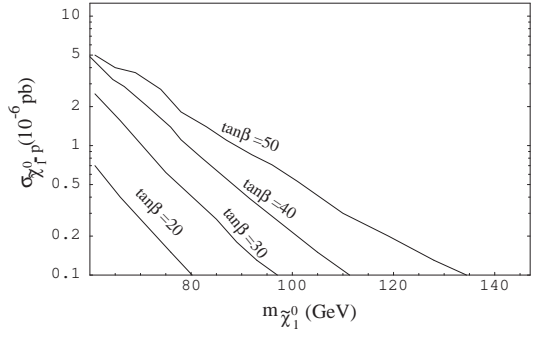

\centerline{ \DESepsf(aads20304050.epsf width 7 cm) }
\bigskip
\bigskip
\caption {\label{fig1}  $(\sigma_{\tilde{\chi}_{1}^{0}-p})_{\rm max}$ for mSUGRA
obtained by varying $A_0$ over the parameter space for  tan${\beta}=20$, 30, 40,
and 50[11]. The relic density constraint, Eq.(3) has been imposed.}
\end{figure}

\begin{figure}[htb]
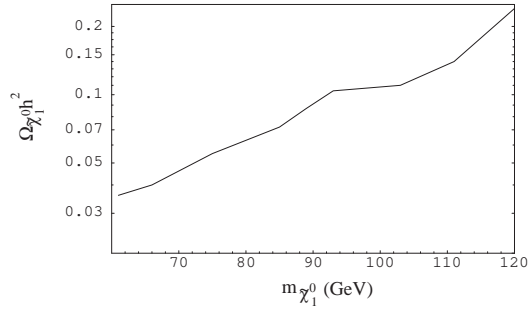

\centerline{ \DESepsf(aads30om.epsf width 7 cm) }
\bigskip
\bigskip
\caption {\label{fig2} $\Omega_{\tilde{\chi}_{1}^{0}}h^{2}$ for mSUGRA when
$(\sigma_{\tilde{\chi}_{1}^{0}-p})$ takes on its  maximum value for
tan${\beta}=30$ [11].}
\end{figure}

\begin{figure}[htb]
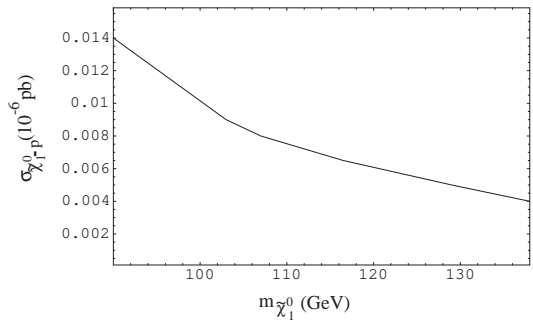

\centerline{ \DESepsf(aadslowlimit.epsf width 7 cm) }
\bigskip
\bigskip
\caption {\label{fig3} $(\sigma_{\tilde{\chi}_{1}^{0}-p})_{\rm min}$ for 
$m_{\tilde{\chi}_{1}^{0}}\leq 140$ GeV for mSUGRA obtained by varying $A_0$  for 
tan${\beta}=3$ with the relic density constraint, Eq.(3)[25].}
\end{figure}
\begin{figure}[htb]
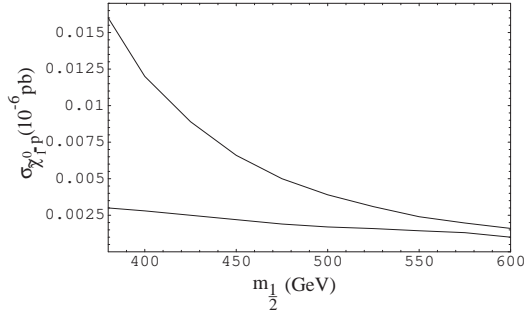

\centerline{ \DESepsf( aadcoan403.epsf width 7 cm) }
\bigskip
\bigskip
\caption {\label{fig4} $(\sigma_{\tilde{\chi}_{1}^{0}-p})$ for mSUGRA in the
coannihilation region for   tan${\beta}=40$ (upper curve) and tan${\beta}=3$
(lower curve), 
$A_0=4 m_{1/2}$, $\mu>0$[25].}
\end{figure}
\begin{figure}[htb]
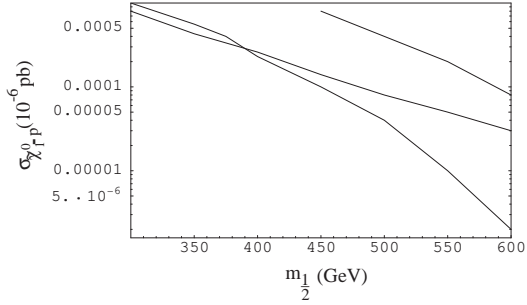

\centerline{ \DESepsf(aadcoan51020.epsf width 7 cm) }
\caption {\label{fig5} $(\sigma_{\tilde{\chi}_{1}^{0}-p})$ for mSUGRA and
$\mu<0$ for (from top to bottom on right)   tan${\beta}$=20, 5 and 10. Note that
for tan${\beta}\geq 10$, the curves terminate at the left due to the
$b\rightarrow s\gamma$ constraint[25].}
\end{figure}
\begin{figure}[htb]
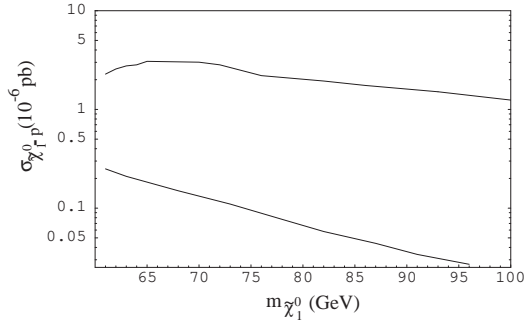

\centerline{ \DESepsf(aads7uni7nonuni.epsf width 7 cm) }
\caption {\label{fig6} Universal mSUGRA (lower curve) and nonuniversal soft
breaking  (upper curve) for tan$\beta=7$, $\mu>0$. The
nonuniversal curve has 
$\delta_3$, $\delta_2$, $\delta_1<0$ $\delta_2>0$[11].}
\end{figure}
\begin{figure}[htb]
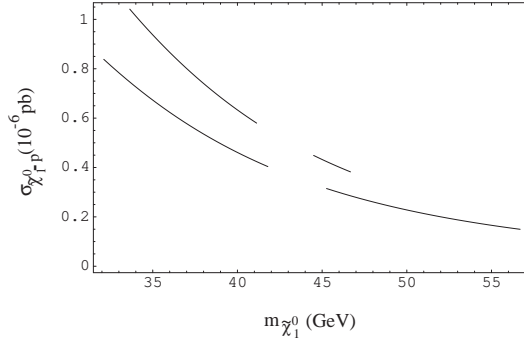

\centerline{ \DESepsf( darktalk20001.epsf width 7 cm) }
\caption {\label{fig7} $\sigma_{\tilde{\chi}_{1}^{0}-p}$  for the D-brane model
for tan$\beta=10$, $m_{3/2}=175$ GeV, $\mu>0$. Upper curve is $\theta_b$=0.5,
lower curve $\theta_b$=0.2[12].}
\end{figure}\begin{figure}[htb]
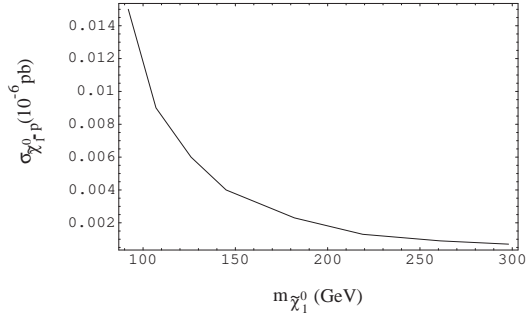

\centerline{ \DESepsf( aadcoandbrane.epsf width 7 cm) }
\caption {\label{fig8} Minimum $\sigma_{\tilde{\chi}_{1}^{0}-p}$  for the
D-brane model for  $m_{3/2}=600$ GeV, $\mu>0$[25].}
\end{figure}

\begin{thebibliography}{99}
\bibitem{com}While the lower bound of Eq.(13) is somewhat lower than other
estimates, it allows us to consider the possibility that not all the dark matter
are neutralinos, i.e. the dark matter might be a mix of neutralinos, machos,
axions etc. Further, the minimum values of $\sigma_{\tilde\chi^0_1-p}$ are not
particularly sensitive to the lower bound $\Omega_{\tilde\chi^0_1}h^2$.
\bibitem{isa} We use the ISAJET sign convention for $\mu$ and $A$.
\bibitem{comm2}If one allows $\delta_i$ to get unusually large, i.e.
$\delta_{1,2}\stackrel{>}{\sim}10$, (as in
\cite{ellis}) or allows the $\tilde\tau_1$ to remain light by setting
$\delta_5\simeq -1$, one can broaden the region of parameter space where the
above cancellation occurs.
\end{thebibliography}
\end{document}